\documentclass[aps,superscriptaddress,twocolumn,showpacs]{revtex4-1}
\pdfoutput=1
\usepackage{color}
\usepackage{amsmath, amsthm, amsfonts}    % need for subequations
\usepackage{amssymb}
\usepackage{mathtools}
\usepackage{mathptmx} 
\usepackage[table]{xcolor}
\usepackage{dsfont}
\usepackage{enumitem} 
\usepackage{appendix}
\usepackage{braket}
\usepackage[skins,theorems]{tcolorbox}
\tcbset{highlight math style={enhanced,
  colframe=red,colback=white,arc=0pt,boxrule=1pt}}
\usepackage{cancel}
\usepackage{empheq}
\usepackage{lipsum}% http://ctan.org/pkg/lipsum
\usepackage{graphicx}% http://ctan.org/pkg/graphicx
\usepackage{tikz}
\usepackage{makecell}
\usepackage[unicode=true,pdfusetitle,
 bookmarks=true,bookmarksnumbered=false,bookmarksopen=false,
 breaklinks=true,pdfborder={0 0 0},backref=false,colorlinks=true,citecolor=blue]{hyperref}
\usepackage{graphicx}   % need for figures
\usepackage[caption=false]{subfig}       % use for side-by-side figures and   to have captions justified and small letter size
\usepackage{bm}            % bold math
\usepackage[normalem]{ulem}  % use for underlining

\newlength\imagewidth
\newlength\imagescale
\def\be{\begin{eqnarray}}
\def\ee{\end{eqnarray}}

\def\u{{\bf u}}

\DeclareUnicodeCharacter{2212}{-}

\definecolor{JOT-color}{named}{blue}
\definecolor{CSF-color}{named}{orange}

\begin{document}
 
%\title{Characterizing cylindrical particles through local measurements of the Stokes parameters} - Jon -
\title{Characterizing cylindrical particles upon local measurements of two Stokes parameters}
%\title{The Stokes Parameter Method for  Characterizing cylindrical particles in far-field}

\author{Jon Lasa-Alonso}
\email{jonqnanolab@gmail.com}
\affiliation{Donostia International Physics Center, Paseo Manuel de Lardizabal 4, 20018 Donostia-San Sebastian, Spain.}
\affiliation{Centro de F\'isica de Materiales, Paseo Manuel de Lardizabal 5, 20018 Donostia-San Sebastian, Spain.}

\author{Iker G\'omez-Viloria}
\affiliation{Centro de F\'isica de Materiales, Paseo Manuel de Lardizabal 5, 20018 Donostia-San Sebastian, Spain.}

\author{Álvaro Nodar}
\affiliation{Centro de F\'isica de Materiales, Paseo Manuel de Lardizabal 5, 20018 Donostia-San Sebastian, Spain.}

\author{Aitzol Garc\'ia-Etxarri}
\affiliation{Donostia International Physics Center, Paseo Manuel de Lardizabal 4, 20018 Donostia-San Sebastian, Spain.}
\affiliation{IKERBASQUE, Basque Foundation for Science, Mar\'ia D\'iaz de Haro 3, 48013 Bilbao, Spain.}

\author{Gabriel Molina-Terriza}
\affiliation{Donostia International Physics Center, Paseo Manuel de Lardizabal 4, 20018 Donostia-San Sebastian, Spain.}
\affiliation{Centro de F\'isica de Materiales, Paseo Manuel de Lardizabal 5, 20018 Donostia-San Sebastian, Spain.}
\affiliation{IKERBASQUE, Basque Foundation for Science, Mar\'ia D\'iaz de Haro 3, 48013 Bilbao, Spain.}

\author{Jorge Olmos-Trigo}
\email{jolmostrigo@gmail.com}
\affiliation{Centro de F\'isica de Materiales, Paseo Manuel de Lardizabal 5, 20018 Donostia-San Sebastian, Spain.}

\begin{abstract}
Researchers routinely characterize optical samples by computing the scattering cross-section. However, the experimental determination of this magnitude requires the measurement and integration of the components of the scattered field in all directions. Here, we propose a method to determine the scattering cross-section and global polarization state of radiation through measurements of two Stokes parameters at an angle of choice in far-field. The method applies to cylindrically symmetric samples whose optical response is well-described by a single multipolar order $j$. Moreover, the formalism is applicable for a wide range of different illuminations, and it only requires the use of a single camera and conventional wave plates. Our findings significantly 
reduce the complexity of routine characterization measurements for cylindrical samples in optical laboratories. 
\end{abstract}

\maketitle

The conservation of energy implies that it cannot be created or destroyed, only transferred or transformed. This principle is essential to understand the behavior of various physical systems, including the scattering of electromagnetic radiation. In a general linear electromagnetic scattering problem, energy is extinguished in two ways: scattering and absorption. Extinction refers to the reduction in the intensity of a light beam as it passes through a medium, scattering is the deflection of light in different directions, and absorption is the process by which a sample consumes energy, usually transforming it into heat. These processes are all governed by energy conservation and can be described quantitatively using cross-sections. The cross-sections can be understood as the area of a sample that is effectively available for light to interact with. As a result, the cross-sections depend on the sample but also on the incident illumination. The conservation of energy is usually expressed in terms of the extinction, $\sigma_{\rm{ext}}$, scattering, $\sigma_{\rm{sca}}$, and absorption cross-sections, $\sigma_{\rm{abs}}$, of the sample through~\cite{bohren2008absorption}:
\begin{equation}  \label{Energy_1}
\sigma_{\rm{ext}} = \sigma_{\rm{sca}} +  \sigma_{\rm{abs}}.
\end{equation}
Thus, measuring the scattering and absorption cross-sections is crucial to understand how energy transfers from light to matter. For example, the optical theorem relates the extinction cross-section of a scatterer with the forward scattering amplitude~\cite{novotny2012principles, evlyukhin2016optical}. Thus, the optical theorem  greatly simplifies the measurement of $\sigma_\text{ext}$ for samples under plane wave illumination. In particular, for lossless samples, the optical theorem  also gives access to the scattering cross-section since in these situations $\sigma_{\rm{ext}} = \sigma_{\rm{sca}}$. However, it is essential to note that the optical theorem  has some limitations: it neglects the complex structure of focused beams, limiting its applicability to plane wave illumination~\cite{newton1976optical,lock1995failure, krasavin2018generalization}. On the other hand, for lossy scatterers, direct measurements of the absorption or scattering cross-sections are required to determine how light is extinguished in the scattering process. Thus, in general, the characterization of linear scatterers requires capturing the extinction, scattering, or absorption cross-sections through alternative means.

One method to determine the absorption cross-section is measuring temperature changes in the sample during the interaction with the incident beam (the Joule effect) ~\cite{baffou2010nanoscale, giannini2011plasmonic}. This method is demanding as the absorption cross-section does not have a universal relationship with temperature, and it is highly dependent on the material, shape, and size of the scatterer~\cite{baffou2013thermo}. An alternative approach is to capture the power scattered by a sample in order to determine its scattering cross-section. However, measuring the scattering cross-section is also challenging as it requires integrating all the scattered field components. In any case, obtaining the extinction, scattering, and absorption cross-sections of samples is essential for many applications spanning different fields of photonics. We can find examples of the pivotal role of scattering and absorption cross-sections in the context of resonant nanoparticles~\cite{luk2010fano, garcia2011strong, kuznetsov2012magnetic, kuznetsov2016optically, luk2015optimum}, optical forces~\cite{nieto2010optical, gomez2012electric, gomez2012negative, olmos2019asymmetry} and torques~\cite{marston1984radiation, nieto2015optical, canaguier2014transverse, xu2019azimuthal}, Kerker conditions~\cite{nieto2011angle, geffrin2012magnetic, fu2013directional, person2013demonstration, liu2014ultra, olmos2020kerker, Resonant, Correlations, olmos2020unveiling, olmos2020optimal,  schmidt2015isotropically, xu2020kerker}, optical anapoles~\cite{miroshnichenko2015nonradiating, luk2017hybrid, feng2017ideal, ho2013anapole, zanganeh2021anapole, parker2020excitation, wei2016excitation, sanz2021multiple}, or surface-enhanced circular dichroism spectroscopy~\cite{tang2010optical, garcia2013surface, ho2017enhancing, hu2019high,graf2019achiral,solomon2018enantiospecific, lasa2020surface,  olmostrigo2023capturing}, among others.

In this work, we present a method to capture the scattering cross-section and the global polarization state of light from local densities of the Stokes parameters. That is, we show that by measuring both intensity $I(r, \theta)$ and degree of circular polarization $V(r, \theta)$ of the scattered  field, at any particular angle of choice $\theta$ in far-field, we can obtain the scattering cross-section and global polarization state of light for cylindrically symmetric samples. Importantly, our proposed method only requires a single camera and a few wave plates placed at a fixed position far from the scatterer. In principle, 
our approach applies when the electric and magnetic optical response of the sample can be fully described by one multipolar order $j$. However, we show that, under the appropriate illumination conditions, all-optical resonances of generic cylindrical particles can be determined through our proposed technique. Therefore, our findings are relevant for any photonic branch dealing with the interaction between electromagnetic waves and cylindrically symmetric samples~\cite{luk2010fano, garcia2011strong, kuznetsov2012magnetic, kuznetsov2016optically, luk2015optimum, nieto2010optical, gomez2012electric, olmos2019asymmetry, gomez2012negative,marston1984radiation, nieto2015optical, canaguier2014transverse, xu2019azimuthal,nieto2011angle, geffrin2012magnetic, fu2013directional, person2013demonstration, liu2014ultra, olmos2020kerker, olmos2020unveiling, olmos2020optimal, schmidt2015isotropically, xu2020kerker,miroshnichenko2015nonradiating, Correlations, Resonant, luk2017hybrid, feng2017ideal, ho2013anapole, zanganeh2021anapole, parker2020excitation, wei2016excitation, sanz2021multiple,tang2010optical, garcia2013surface, ho2017enhancing, hu2019high,graf2019achiral,solomon2018enantiospecific, lasa2020surface, olmostrigo2023capturing}.

It is a well-known result from electrodynamics that the electromagnetic fields radiated by an arbitrary linear scatterer can be expanded in terms of multipolar fields. Following Jackson's notation in its third edition~\cite{jackson1999electrodynamics}, one can write:
\begin{align}
    \nonumber
    &\mathbf{E}(\mathbf{r}) = Z\sum_{j,m} \left[ ia_E(j,m)\mathbf{N}^h_{j m}(\mathbf{r}) + a_M(j,m)\mathbf{M}^h_{j m}(\mathbf{r}) \right],\\
    &iZ\mathbf{H}(\mathbf{r}) = \frac{1}{k}\nabla\times\mathbf{E}(\mathbf{r}).
    \label{RadJackson}
\end{align}
Here $\mathbf{E}(\mathbf{r})$ and $\mathbf{H}(\mathbf{r})$ are the scattered electric and magnetic fields, respectively; $Z = \sqrt{\mu/ \epsilon}$ is the impedance of the medium in which the scatterer is embedded, $\mu$ and $\epsilon$ being the magnetic permeability and electric permittivity, respectively, and $k$ is the modulus of the wavevector associated to the radiation. Moreover,  $a_E(j,m)$ and $a_M(j,m)$ are the electric and magnetic coefficients; $\mathbf{M}^h_{j m}(\mathbf{r}) = h^{\scriptscriptstyle (1)}_j(kr)\mathbf{X}_{j m}(\theta,\varphi)$ and $\mathbf{N}^h_{j m}(\mathbf{r}) = k^{-1}\nabla \times \mathbf{M}^h_{j m}(\mathbf{r})$ are Hansen multipoles, with $h^{\scriptscriptstyle (1)}_j (kr)$ the spherical Hankel functions of the first kind and $\mathbf{X}_{j m}(\theta, \varphi)$ the vector spherical harmonics as defined in Ref.~\cite{jackson1999electrodynamics}. Here $j$ and $m$ denote the multipolar order and the eigenvalue of the total angular momentum in the $z$ direction, respectively.

The description of the scattered fields given in Eq.~\eqref{RadJackson} is general and applies to an extensive range of different radiating systems. However, for the aim of this work, we will focus on cylindrical samples illuminated by beams with a well-defined total angular momentum $m$. In particular, we will tackle scatterers that may be well-described by a single multipole order $j$ in Eq.~\eqref{RadJackson}. This implies that the response of the sample is given by both electric and magnetic scattering coefficients with $j = 1$ (dipolar), $j = 2$ (quadrupolar), $j = 3$ (octupolar)... etc. On the other hand, when particularizing to cylindrically symmetric samples illuminated by beams with well-defined total angular momentum, we are also fixing the value of parameter $m$ in the sum given by Eq.~\eqref{RadJackson}. The physical reason behind this phenomenon resides in the fact that, because of their symmetry, cylindrical samples preserve the $z$ component of the total angular momentum upon light-matter interactions. Hence, it is direct to check that when considering this specific type of scatterers, we can get rid of the summation over both $j$ and $m$ indices in the expression of the scattered field given by Eq. \eqref{RadJackson}. 

In addition, if we switch to the helicity basis, i.e. $\mathbf{E}^{\pm}(\mathbf{r}) =  \left[\mathbf{E}(\mathbf{r}) \pm iZ \mathbf{H}(\mathbf{r}) \right] / 2$, we are led to the following compact expression of the scattered fields for this particular type of scatterers:
\begin{equation} \label{E_sigma}
    \mathbf{E}^{\pm}(\mathbf{r}) = a_{\pm}(j, m)\left(\frac{\mathbf{N}^h_{j m}(\mathbf{r}) \pm \mathbf{M}^h_{j m}(\mathbf{r})}{\sqrt{2}}\right).
\end{equation}
Here, $\mathbf{E}^{\pm}(\mathbf{r})$ are proportional to the monochromatic Riemann- Silberstein vectors~\cite{Birula1, RoleRSBirula} and  $a_{\pm}(j, m) = Z[ia_E(j, m) \pm a_M(j, m)]/ \sqrt{2}$. From now on, the $(j, m)$ multipolar dependence will be assumed in the scattering coefficients $a_{\pm}(j, m)$, namely,  $a_{\pm} \equiv a_{\pm}(j, m)$ . Our interest in choosing the helicity basis lies in the fact that, in the far-field, the local fields are related to the Stokes parameters. In this regard, fundamental magnitudes such as the total scattering cross-section, proportional to the $I$ parameter, or the helicity expectation value, constructed as the ratio $V/I$, can be obtained from the far-field expressions of the scattered electromagnetic field in the helicity basis~\cite{crichton2000measurable}. In the far-field, we can write Eq.~\eqref{E_sigma} as~\cite{Carrascal}:
\begin{equation}
    \label{FieldsFar}
    \mathbf{E}^{\pm}_{\text{far}}(\mathbf{r}) = -a_{\pm} \frac{e^{ikr}}{kr}\frac{(-i)^{j+1}}{\sqrt{j(j+1)}}\left(\frac{\boldsymbol{\xi}_{j m}(\theta,\varphi) \pm i\boldsymbol{\eta}_{j m}(\theta, \varphi)}{\sqrt{2}}\right),
\end{equation}
where we have defined the complex vector functions $\boldsymbol{\xi}_{j m}(\theta, \varphi) = r\nabla Y_{j m}(\theta, \varphi)$ and $\boldsymbol{\eta}_{j m}(\theta, \varphi) = \hat{r} \times \boldsymbol{\xi}_{j m}(\theta, \varphi)$, with $Y_{j m}(\theta, \varphi)$ the scalar spherical harmonics. 

The measurable total scattered intensity $\langle I \rangle_{\rm{sca}}$ and total degree of circular polarization $\langle V \rangle_{\rm{sca}}$ in the far-field can be written as:
\begin{align}
    \label{Ufar}
    \langle I \rangle_{\rm{sca}} = (kr)^2\int~I_\text{far}(r,\theta)~d\Omega, \\
    \label{Lfar}
    \langle V \rangle_{\rm{sca}} = (kr)^2 \int~V_\text{far}(r,\theta)~d\Omega,
\end{align}
where $I_\text{far}(r,\theta) = (|\mathbf{E}^{+}_\text{far}(\mathbf{r})|^2 + |\mathbf{E}^{-}_\text{far}(\mathbf{r})|^2)/ |E_0|^2$ and $V_\text{far}(r, \theta) = (|\mathbf{E}^{+}_\text{far}(\mathbf{r})|^2 - |\mathbf{E}^{-}_\text{far}(\mathbf{r})|^2)/|E_0|^2$ are, the local scattered intensity and the local degree of circular polarization in the far-field, respectively. Note that $E_0$ is the amplitude of the incident electric field and, as a result, $I_\text{far}(r,\theta)$ and $V_\text{far}(r,\theta)$ are dimensionless.  For a sample whose response is well-described by a single multipolar order $j$, the explicit computation of the integrals given in Eq.~\eqref{Ufar} and Eq.~\eqref{Lfar} yields
\begin{align} \label{IVfar2}
    \langle I \rangle_{\rm{sca}} = \frac{|a_+|^2 + |a_-|^2}{|E_0|^2}, &&
    \langle V \rangle_{\rm{sca}} =\frac{|a_+|^2 - |a_-|^2}{|E_0|^2} .
\end{align}
The results given in Eq.~\eqref{IVfar2} have already been reported previously~\cite{jackson1999electrodynamics}. However, their connection to the local intensity and local degree of circular polarization densities for cylindrical samples has not been studied in detail.

Let us explicitly compute the intensity and degree of circular polarization densities in the far-field. For cylindrical samples well-described by a single multipolar order $j$, $\langle I \rangle_{\rm{sca}}$ and $\langle V \rangle_{\rm{sca}}$ are determined by the following fundamental relation (see Appendix \ref{DerEq48}):
\begin{align}
    \label{Relation}
    \begin{pmatrix}
    \langle I \rangle_{\rm{sca}} \\
    \langle V \rangle_{\rm{sca}}
    \end{pmatrix} = 
     \frac{(kr)^2j(j + 1)}{f_{j m}^2(\theta) - g_{j m}^2(\theta)}
    \begin{pmatrix}
    f_{j m}(\theta) & g_{j m}(\theta)\\
    g_{j m}(\theta) & f_{j m}(\theta)
    \end{pmatrix}
    \begin{pmatrix}
    I_\text{far}(r, \theta)\\
    V_\text{far}(r, \theta) 
    \end{pmatrix},
\end{align}
with the real scalar functions $f_{j m}(\theta) = |\boldsymbol{\xi}_{j m}(\theta,\varphi)|^2$ and $g_{j m}(\theta) = \text{Im}[ \boldsymbol{\xi}^*_{j m}(\theta,\varphi)\cdot \boldsymbol{\eta}_{j m}(\theta,\varphi)]$.   The expression given by Eq.~\eqref{Relation} constitutes the main result of this work. It indicates that, for cylindrical samples well-described by a single multipolar order $j$, fundamental quantities such as the integrated scattered intensity or total degree of circular polarization are related to their local densities in the far-field computed at any particular scattering angle $\theta$. That is, by measuring $I_\text{far}(r, \theta)$ and $V_\text{far}(r, \theta)$ in one direction  far from the illuminated sample, we can straightforwardly obtain $\langle I \rangle_{\rm{sca}}$ and $\langle V \rangle_{\rm{sca}}$ by applying a matrix that, in turn, can be easily evaluated once $j$ and $m$ are fixed~\cite{jackson1999electrodynamics}. It is important to note that the matrix elements do not depend on the optical response of the cylindrical scatterers (see Appendix \ref{fjm_g_jm}).

To get a deeper insight into the generality of Eq.~\eqref{Relation}, let us now introduce one of the most illustrative examples in the nanophotonics community: cylindrically symmetric dipolar objects under the illumination of a circularly-polarized plane-wave~\cite{luk2010fano, garcia2011strong, kuznetsov2012magnetic, kuznetsov2016optically, luk2015optimum, nieto2010optical, gomez2012electric, gomez2012negative, olmos2019asymmetry, marston1984radiation, nieto2015optical, canaguier2014transverse, xu2019azimuthal,nieto2011angle, geffrin2012magnetic, fu2013directional, person2013demonstration, liu2014ultra, olmos2020kerker, olmos2020optimal,olmos2020unveiling,  schmidt2015isotropically, xu2020kerker,miroshnichenko2015nonradiating, Correlations, Resonant, luk2017hybrid, feng2017ideal, ho2013anapole, zanganeh2021anapole, parker2020excitation, wei2016excitation, sanz2021multiple,tang2010optical, garcia2013surface, ho2017enhancing, hu2019high,graf2019achiral,solomon2018enantiospecific, lasa2020surface, olmostrigo2023capturing}. This particular physical system fixes $j = 1$ (electric and magnetic dipolar response) and $m = p$, where $p = \pm 1$ is the helicity of the incident beam. In this scenario, it can be shown (see Appendix~\ref{dipolos}) that Eq.~\eqref{Relation} reduces to the following expression:
\begin{align}
    \label{Relation_dipolar}
    \begin{pmatrix}
    \langle I \rangle_\text{sca} \\
    \langle V \rangle_\text{sca}
    \end{pmatrix} = 
     \frac{16 \pi (kr)^2}{3 \sin^4 \theta}
    \begin{pmatrix}
    1 + \cos^2 \theta & -2 p \cos \theta\\
    -2 p \cos \theta & 1 + \cos^2 \theta
    \end{pmatrix}
    \begin{pmatrix}
    I_\text{far}(r, \theta)\\
    V_\text{far}(r, \theta) 
    \end{pmatrix}.
\end{align}
Equation~\eqref{Relation_dipolar} can  be used as a roadmap to infer global properties of the scattered field by cylindrical symmetry dipolar objects, such as $\langle I \rangle_\text{sca}$ and $\langle V \rangle_\text{sca}$, from a single measurement of local Stokes parameters. Then, from $\langle I \rangle_\text{sca}$ and $\langle V \rangle_\text{sca}$, we can construct widely employed magnitudes such as the scattering cross-section, $\sigma_\text{sca}$, and the helicity expectation value, $\langle \Lambda \rangle$~\cite{Correlations, olmos2020unveiling}. More explicitly, these magnitudes can be fixed through the following relations:
\begin{align} \label{expression}
k^2\sigma_\text{sca} =  \langle I \rangle_\text{sca}, && \langle \Lambda \rangle = \frac{\langle V \rangle_\text{sca}}{\langle I \rangle_\text{sca}}.
\end{align}
Equation~\eqref{expression} shows that one can compute the scattering cross-sections and the helicity expectation value from local Stokes parameters, as we previously anticipated. Note that the absorption cross-section, $\sigma_{\rm{abs}}$, can be straightforwardly obtained if the extinction cross-section, $\sigma_{\rm{ext}}$, is  captured by other means (see Eq.~\eqref{Energy_1}).

At this stage, we have all the necessary components to demonstrate the power of our novel method. In this regard, we would like to highlight one of the most significant contributions to the field of high-refractive-index (HRI), ``Magnetic Light" by Arseniy Kuznetsov et al.~\cite{kuznetsov2012magnetic}. In this work, the authors showed the experimental dark-field scattering spectra of spherical Silicon nanoparticles. Note that Silicon has losses in this wavelength range. Furthermore, they presented the theoretical scattering and extinction spectra calculated by Mie theory for spherical Silicon nanoparticles of different sizes in free space within the visible spectral range. 

\begin{figure}[t!]
    \centering
\includegraphics[width=\columnwidth]{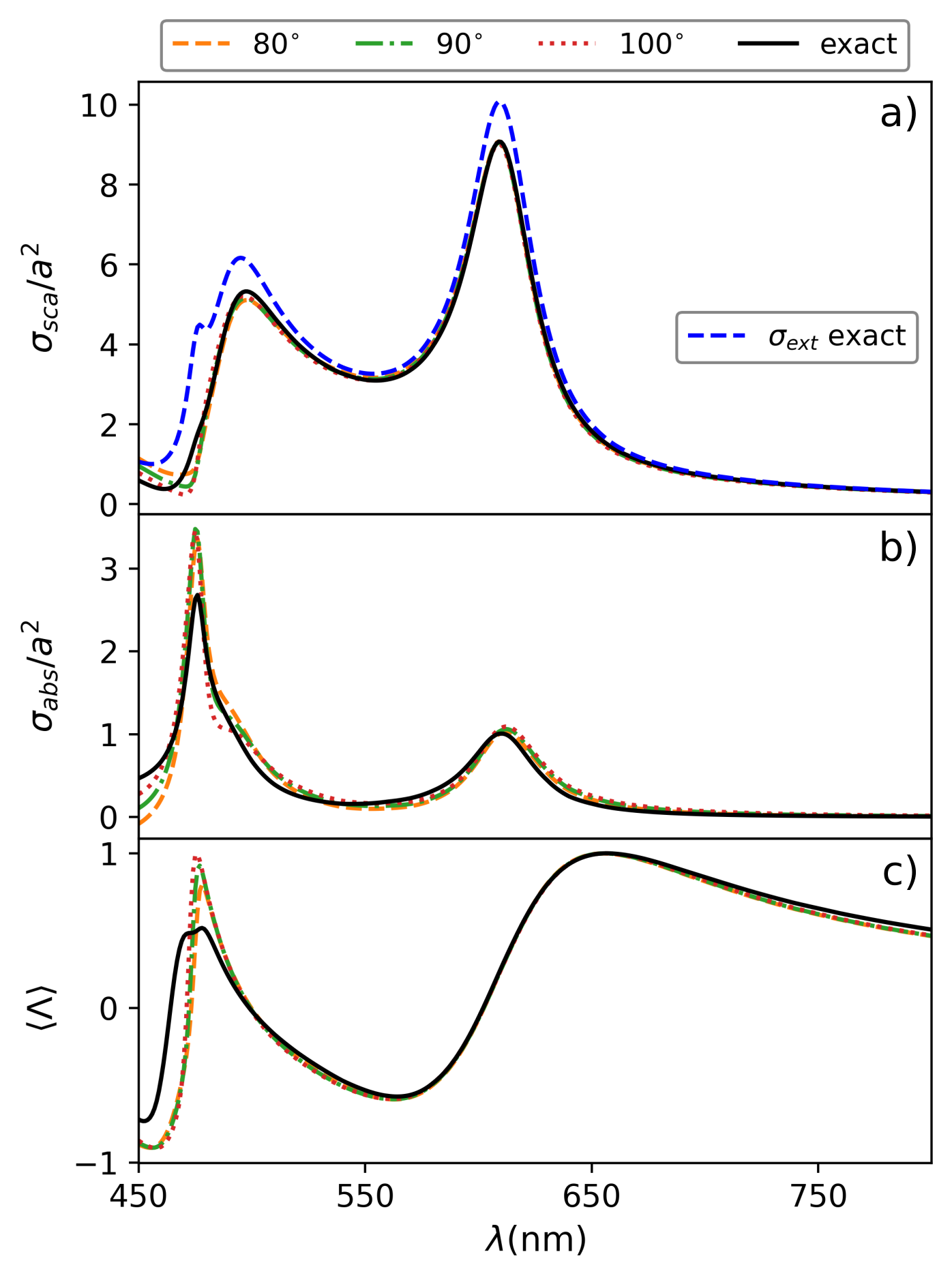}
\caption{a) Scattering efficiency, $\sigma_\text{sca}/a^2$ (solid black), and extinction efficiency, $\sigma_\text{ext}/ a^2$ (dashed blue), calculated from Mie theory for a Silicon nanosphere with radius $a = 75$ nm in the visible spectral range. The scattering cross-sections are calculated from Eqs.~\eqref{Relation_dipolar}-\eqref{expression} by computing $I_\text{far}(\theta,r)$ and $V_\text{far}(\theta,r)$ at $\theta = 80^{\circ}$ (dashed orange), $\theta = 90^{\circ}$ (dashed green) and $\theta = 100^{\circ}$ (dashed red). b) Absorption efficiency, $\sigma_\text{abs}/a^2$ (solid black), calculated from Mie theory. Absorption efficiencies calculated from $\sigma_\text{ext}- \sigma_\text{sca} (\theta)$ for $\theta = 80^{\circ}$ (dashed orange), $\theta = 90^{\circ}$ (dashed green), $\theta = 100^{\circ}$ (dashed red). c) Helicity expectation value, $\langle \Lambda \rangle$, calculated from Mie theory (solid black). The helicity expectation value at different angles is calculated from Eqs.~\eqref{Relation_dipolar}-\eqref{expression} by computing $I(\theta,r)$ and $V(\theta,r)$ at $\theta = 80^{\circ}$ (dashed orange), $\theta = 90^{\circ}$ (dashed green) and $\theta = 100^{\circ}$ (dashed red).}
    \label{rusos}
\end{figure}

In Fig.~\ref{rusos}a-c, we depict several magnitudes for a Silicon nanosphere with radius $a = 75$ nm (Fig. 3c in Ref.~\cite{kuznetsov2012magnetic}) under the illumination of a circularly polarized plane-wave computed both from Mie theory~\cite{mie1908beitrage, bohren2008absorption} and through our novel method, summarized in Eqs.~\eqref{Relation_dipolar}-\eqref{expression} for dipolar objects. Let us just remark that the results reported for different angles $\theta$ are associated with measurements of the local Stokes parameters with a single camera placed far from the sample. In Fig.~\ref{rusos}a), we show the scattering efficiency, $\sigma_\text{sca}/a^2$ (solid black), and the extinction efficiency, $\sigma_\text{ext}/a^2$ (dashed blue), calculated from Mie theory. Moreover, we show the scattering efficiency calculated from Eqs.~\eqref{Relation_dipolar}-\eqref{expression} for different scattering angles, namely, $\theta = 80^{\circ}$ (dashed orange), $\theta = 90^{\circ}$ (dashed green) and $\theta = 100^{\circ}$ (dashed red). The agreement is total in the broadband wavelength interval of $475$ nm $< \lambda <$ $800$ nm for every considered scattering angle. That is, when the spherical particle can be fully described by electric and magnetic dipolar modes. This agreement confirms that, indeed, global magnitudes such as the scattering cross-section can be computed from local measurements of the $I$ and $V$ Stokes parameters. 

In Fig.~\ref{rusos}b), we show the absorption efficiency calculated from Mie theory and from the relation $\sigma_{\rm{abs}}(\theta)/a^2 = \sigma_{\rm{ext}}/a^2 - \sigma_{\rm{sca}}(\theta)/a^2$. There is a reasonable agreement within the interval $475$ nm $< \lambda <$ $800$ nm. That is, when electric and magnetic modes can fully describe the electromagnetic response of the spherical nanoparticle. This result indicates that, under plane-wave illumination, one can obtain the absorption cross-section without measuring temperature changes in spherical dipolar particles during the interaction time. Note that this  result holds as well for plasmonic spherical dipolar particles.

Finally, in Fig.~\ref{rusos}c), we depict the helicity expectation value computed from both Mie theory and our novel approach. Note that we also capture $\langle \Lambda \rangle$ with almost no error in the interval $475$ nm $< \lambda <$ $800$ nm. Furthermore, we also show that it is possible to identify the first Kerker condition~\cite{kerker1983electromagnetic,geffrin2012magnetic, fu2013directional, person2013demonstration}, at which helicity is preserved~\cite{fernandez2013electromagnetic, olmos2019enhanced, olmos2022helicity} upon scattering, $\Lambda \sim 1$ at
$\lambda \sim 660$ nm. 

Hitherto, we have shown that the scattering cross-section and the expected value of helicity of dipolar particles under plane-wave illumination can be determined through the relations specified in Eqs.~\eqref{Relation_dipolar}-\eqref{expression}. In this regard, we have proved that the absorption cross-section can also be computed (via extinction) from the scattering cross-section for these type of samples. In other words, we have shown that our method permits the characterization of small cylindrical particles under plane-wave illumination. In the forthcoming, we show that our approach also allows the characterization of larger cylindrically symmetric particles beyond the dipolar regime. For that aim, we take advantage of the fact that the relation specified by Eq.~\eqref{Relation} works under a wider set of illumination conditions. In particular, we anticipate that tightly-focused Laguerre-Gaussian (LG) beams with well-defined helicity, $p$, and total angular momentum $m = \ell + p$, where $\ell$ is the orbital angular momentum, can be employed for this task.

\begin{figure}[t]
    \centering
\includegraphics[width=\columnwidth]{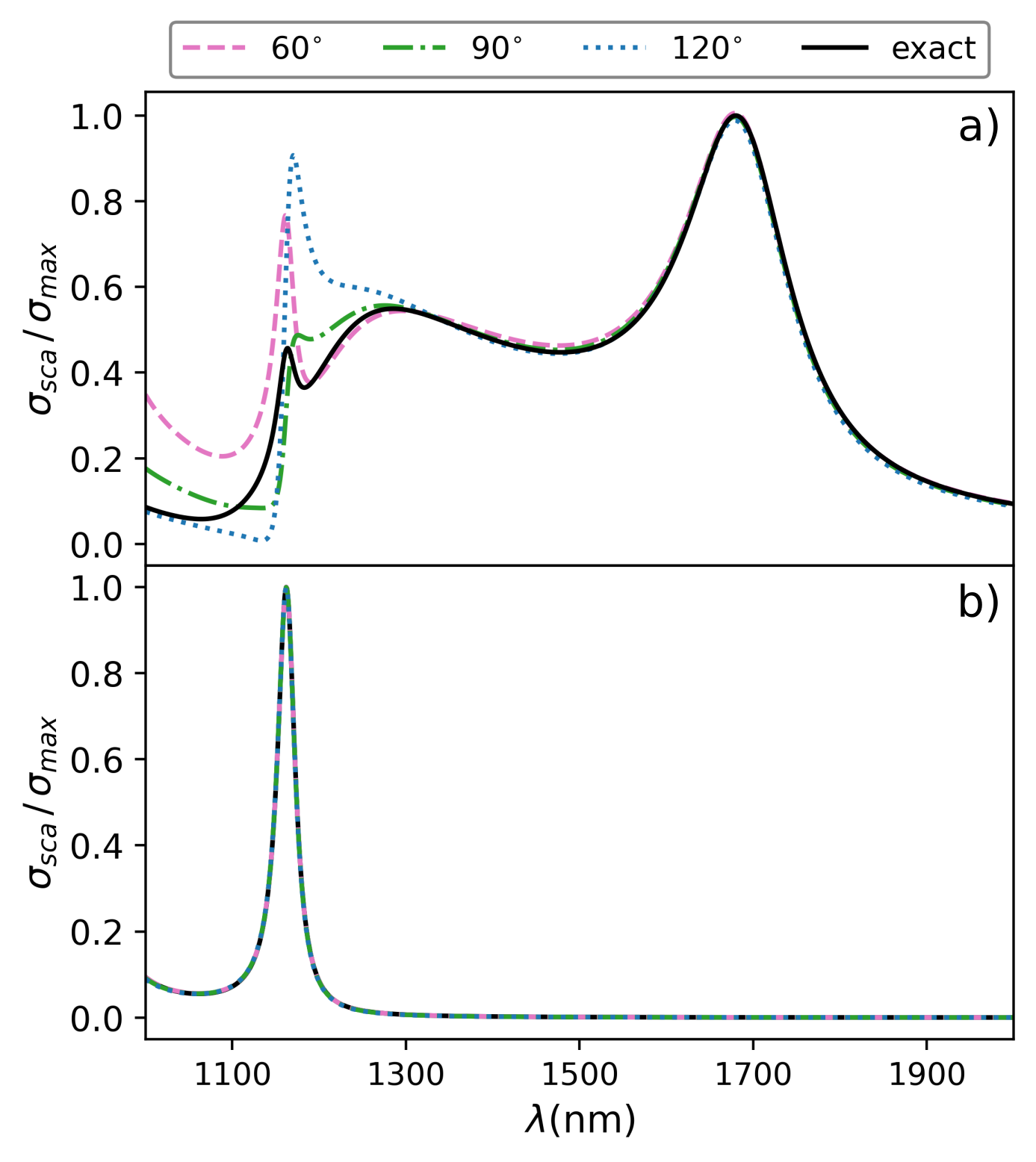}
\caption{Normalized scattering cross-section, $\sigma_\text{sca}/\sigma_\text{max}$, for a Silicon nanosphere of radius $a = 230$ nm calculated from the Generalized Lorentz-Mie theory (solid black) and at $\theta = 60^\circ$ (dashed pink), $\theta = 90^\circ$ (dashed green), $\theta = 120^\circ$ (dotted blue). $\lambda$ represents the wavelength of the incident field. a) The incident field is a tightly-focused LG beam with total angular momentum $m = 1$ and helicity $p = 1$. b) The incident field is a tightly-focused LG beam with total angular momentum $m = 2$ and helicity $p = 1$. In both cases, the numerical aperture is $\rm{NA} = 0.9$. }
    \label{mole}
\end{figure}

Let us now show all of this. In particular, we would like to drive attention to another seminal contribution in the field of HRI nanoparticles: ``Strong magnetic response of submicron Silicon particles in the infrared" by Garc{\'i}a-Etxarri et al~\cite{garcia2011strong}. In this work, the authors showed the scattering spectra calculated from Mie theory for a spherical Silicon nanoparticle of radius $a = 230$ nm in the near-infrared. Note that Silicon has no losses in this wavelength range. We first excite such Silicon nanosphere with a tightly-focused LG beam with $\ell = 0$ and $p = 1$~\cite{gouesbet1985scattering, gouesbet2011generalized}. The focusing process is considered through an aplanatic lens of numerical aperture $\rm{NA} = 0.9$. As the total angular momentum for this type of beam is $m = 1$, we may still employ the relation given by Eq. \eqref{Relation_dipolar}. Consequently, our method should facilitate the identification of dipolar resonances. As we can infer from Fig.~\ref{mole}a), the scattering cross-section is fully captured from local measurements at either $\theta = 60 ^{\circ}$, $\theta = 90^{\circ}$ or $\theta = 120^{\circ}$ in the wavelength interval of $1300$ nm $< \lambda < 2000$ nm. That is, when the scattering can be fully described with electric and magnetic dipoles. However, due to the contribution of the quadrupole, the scattering cross-section is ill-captured for $\lambda < 1300$ nm as the scattering cannot be fully described by a single multipolar order $j$. Note that the disagreement in the local measurements at $\lambda \sim 1300$ nm for different scattering angles is useful to identify the regime at which the sample ceases to be dipolar.

At this point, as our method works for a wider range of illuminating fields, we may change the incident beam and excite the scatterer with a tightly-focused LG of $\ell = 1$ and $p = 1$ to fully capture the quadrupolar resonance. %while not exciting the dipolar contribution.
The result is depicted in Fig.~\ref{mole}b). Such a remarkable outcome stems from the fact that the illuminating beam has a total angular momentum $m = 2$ and, thus, it cannot excite any dipolar ($j = 1$) resonances within the scatterer. It is a well-known result that incident fields with total angular momentum $m$ cannot excite multipolar resonances of order $j < |m|$ in cylindrically symmetric scatterers~\cite{zambrana2012excitation}. As a result, the underlying dipolar resonances are cleared out, and our method remains valid for the characterization of the particle in the range $\lambda < 1300$ nm. Of course, when exciting a quadrupolar mode ($j = 2$) with an LG beam with total angular momentum $m = 2$, the matrix given by Eq. \eqref{Relation} has to be re-evaluated. In this case,  the functions that have to be employed are $f_{22}(\theta) = (15/8\pi)\sin^2\theta(1 + \cos^2\theta)$ and $g_{22}(\theta) = -(15/4\pi)\sin^2\theta\cos\theta$. Note that this procedure is recursive, i.e., an LG beam of $\ell = 2$ and $p = 1$ can be employed to characterize octupolar ($j = 3$) spectral regions and so on. Also, the numerical aperture employed in Fig.~\ref{mole}a) and \ref{mole}b) is the same. This implies that the characterization method can be implemented within a single set up in which the orbital angular momentum of the incident beam is switchable.

In conclusion, we have presented a novel method to capture the scattering cross-section and global polarization state of radiation through local measurements of the Stokes parameters at an angle of choice in the far-field. This novel approach requires a single camera and conventional wave plates, significantly reducing the complexity of routine characterization measurements for cylindrically symmetric samples in optical laboratories.
Moreover, the method is applicable under a wide range of illumination conditions, which implies that it can be used in a wide range of experimental settings. For instance, we have shown its applicability for plane waves and LG beams of different orbital angular momenta. In particular, we have shown that the use of tightly-focused LG beams permits the spectral characterization of multipolar cylindrical scatterers. Experimentally, the construction of LG beams with different $\ell$ values can be carried out through the use of a Spatial Light Modulator, which may easily be integrated in an optical set up. Thus, our characterization method is completely feasible with the instruments currently available.

Overall, our findings have important implications for the field of optics. By simplifying and streamlining the characterization process, our method could enable researchers to more accurately and efficiently study a wide range of cylindrically symmetric samples in a plethora of photonic branches.

\section*{Acknowledgments}
J.O.T. acknowledges support from the Juan de la Cierva fellowship No. FJC2021-047090-I of  MCIN/AEI
/10.13039/501100011033 and NextGenerationEU/PRTR. J.O.T, J.L.A., and A.G.E acknowledge support from Project No. PID2019-109905GA-C22 of the Spanish Ministerio de Ciencia, Innovación y Universidades (MICIU). We received funding from the IKUR Strategy under the collaboration agreement between Ikerbasque Foundation and DIPC on behalf of the Department of Education of the Basque Government, as well the Basque Government Elkartek program (KK-2021/00082) and the Centros Severo Ochoa AEI/CEX2018-000867-S from the Spanish Ministerio de Ciencia e Innovación. A.N. acknowledges the financial support from the Spanish Ministry of Science and Innovation and the Spanish government agency of research MCIN/AEI/10.13039/501100011033 through Project Ref. No. PID2019-107432GB-I00 and from the Department of Education, Research and Universities of the Basque Government through Project Ref. No. IT1526-22. G.M.T received funding from the IKUR Strategy under the collaboration agreement between Ikerbasque Foundation and DIPC/MPC on behalf of the Department of Education of the Basque Government.

%\JOT{Jorge: Creo que hace falta poner md, ed, mq al lado de las resonancias de las figura 1a y 2a, 2b. Ayuda a leerlo.}
%\JOT{Revisar citas, seguro que se pueden poner el doble. No sé si es necesario.}

\clearpage
\appendix

\section{Response of a cylindrically symmetric object well-described by a single multipolar order $j$} \label{DerEq48}

In this Appendix, we show how to derive the fundamental relation expressed in Eq. \eqref{Relation} starting from the fields in Eq. \eqref{FieldsFar}. In the derivation we omit the explicit dependence of the fields on the $(r,\theta,\phi)$ variables. By taking the modulus squared of the fields expressed in Eq. \eqref{FieldsFar}, we arrive to:
\begin{equation}
    \label{Der1}
    |\mathbf{E}_\text{far}^{\pm}|^2 = |a_\pm|^2\left(\frac{|\boldsymbol{\xi}_{jm}|^2 \mp \text{Im}\left[\boldsymbol{\xi}_{jm}^*\cdot\boldsymbol{\eta}_{jm}\right]}{(kr)^2 j(j+1)}\right),
\end{equation}
where we have used that $|\boldsymbol{\eta}_{jm}|^2 = |\boldsymbol{\xi}_{jm}|^2$. Normalizing the expression by the modulus of the amplitude of the incident electric field, $|E_0|^2$, and taking the sum and difference of the helicity components expressed in Eq. \eqref{Der1} we arrive to:
\begin{align}
    \label{Der2}
    I_\text{far} &= \frac{1}{(kr^2)j(j+1)}\left[ f_{jm} \langle I \rangle_\text{sca} - g_{jm} \langle V \rangle_\text{sca} \right]\\
    \label{Der3}
    V_\text{far} &= \frac{1}{(kr^2)j(j+1)}\left[ -g_{jm} \langle I \rangle_\text{sca} + f_{jm} \langle V \rangle_\text{sca} \right], 
\end{align}
where $I_\text{far} = (|\mathbf{E}^+_\text{far}|^2 + |\mathbf{E}^-_\text{far}|^2)/|E_0|^2$ and $V_\text{far} = (|\mathbf{E}^+_\text{far}|^2 - |\mathbf{E}^-_\text{far}|^2)/|E_0|^2$ are the local Stokes parameters measured fram from the scatterer. Also, we have defined the real scalar functions $f_{jm} = |\boldsymbol{\xi}_{jm}|^2$ and $g_{jm} = \text{Im}[\boldsymbol{\xi}_{jm}^*\cdot\boldsymbol{\eta}_{jm}]$. Finally, we have employed Eq. \eqref{IVfar2} of the main text to express the integrated Stokes parameters $\langle I \rangle_\text{sca}$ and $\langle V \rangle_\text{sca}$.

Note that the relation expressed in Eqs. \eqref{Der2}-\eqref{Der3} can be compactly written in matrix notation as:
\begin{equation}
    \begin{pmatrix}
    I_\text{far}\\
    V_\text{far}
    \end{pmatrix}
    =
    \frac{1}{(kr^2)j(j+1)}
    \begin{pmatrix}
    f_{jm} && -g_{jm}\\
    -g_{jm} && f_{jm}
    \end{pmatrix}
    \begin{pmatrix}
    \langle I \rangle_\text{sca}\\
    \langle V \rangle_\text{sca}
    \end{pmatrix}.
\end{equation}
By taking the inverse of the relation expressed above we finally obtain:
\begin{equation}
        \begin{pmatrix}
    \langle I \rangle_\text{sca}\\
    \langle V \rangle_\text{sca}
    \end{pmatrix}
    =
    \frac{(kr^2)j(j+1)}{f_{jm}^2 - g_{jm}^2}
    \begin{pmatrix}
    f_{jm} && g_{jm}\\
    g_{jm} && f_{jm}
    \end{pmatrix}
    \begin{pmatrix}
    I_\text{far}\\
    V_\text{far}
    \end{pmatrix},
\end{equation}
which is exactly the expression given in Eq. \eqref{Relation} of the main text.

\section{General form of $f_{jm}(\theta)$ and $g_{jm}(\theta)$} \label{fjm_g_jm}

The general form of $f_{jm}(\theta)$ and $g_{jm}(\theta)$ functions is:
\begin{equation}
    f_{jm}(\theta) = C^2_{jm}\left\{ \sin^2\theta \left[\frac{\partial P_j^m(\cos\theta)}{\partial (\cos\theta)}\right]^2 + \frac{m^2}{\sin^2\theta} \left[ P_j^m(\cos\theta) \right]^2 \right\}
\end{equation}
and
\begin{equation}
    g_{jm}(\theta) = 2C_{jm}^2\left\{ m~P_j^m(\cos\theta) \frac{\partial P_j^m(\cos\theta)}{\partial(\cos\theta)} \right\},
\end{equation}
where $P_j^m(\cos\theta)$ are the associated Legendre polynomials. Also, the coefficients $C_{jm}$ are defined as:
\begin{equation}
    C_{jm} = \sqrt{\frac{2j +1}{4\pi}\frac{(j-m)!}{(j+m)!}}.
\end{equation}

\section{Response of a cylindrically symmetric dipolar object} \label{dipolos}
In this Appendix, we show the steps to arrive from Eq.~\eqref{Relation} to Eq.~\eqref{Relation_dipolar}. To start with, let us explicitly write the functions $\boldsymbol{\xi}_{\ell m}(\theta, \varphi)$ and $\boldsymbol{\eta}_{\ell m}(\theta, \varphi)$ for $\ell = 1$. We also fix that the total angular momentum is $m = p =\pm 1$. Following Jackson's book in its third edition~\cite{jackson1999electrodynamics}, we can write
\be
\boldsymbol{\xi}_{1 m}(\theta, \varphi) &=&  -\frac{1}{2} \sqrt{\frac{3}{2 \pi}} \left( \cos \theta e^{i m \varphi} \hat{\u}_{\theta} + i m e^{i m \varphi} \hat{\u}_{\varphi} \right)\\
\boldsymbol{\eta}_{1 m}(\theta, \varphi) &=&  -\frac{1}{2} \sqrt{\frac{3}{2 \pi}} \left( \cos \theta e^{i m \varphi} \hat{\u}_{\theta} - i m e^{i m \varphi}\hat{\u}_{\varphi} \right)
\ee 
Now, let us compute $f_{1 m}(\theta) = |\boldsymbol{\xi}_{1 m}(\theta,\varphi)|^2$ and $g_{1 m}(\theta) = \text{Im}\left[ \boldsymbol{\xi}^*_{1 m}(\theta,\varphi)\cdot \boldsymbol{\eta}_{1 m}(\theta,\varphi)\right]$. After some algebra, we get 
\begin{align} \label{f_g_dipolares}
f_{1 m}(\theta) =  \frac{3}{8 \pi}  \left( 1 + \cos^2 \theta \right), &&
g_{1 m}(\theta) =  -\frac{3}{4 \pi} m \cos \theta.    
\end{align}
Now, by inserting Eq.~\eqref{f_g_dipolares} into Eq.~\eqref{Relation}, it is straightforward to arrive to
\begin{align}
    \label{Relation_dipolar_appendix}
    \begin{pmatrix}
    \langle I \rangle_\text{sca} \\
    \langle V \rangle_\text{sca}
    \end{pmatrix} = 
     \frac{16 \pi (kr)^2}{3 \sin^4 \theta}
    \begin{pmatrix}
    1 + \cos^2 \theta & -2m \cos \theta\\
    -2 m \cos \theta & 1 + \cos^2 \theta
    \end{pmatrix}
    \begin{pmatrix}
    I_\text{far}(r, \theta)\\
    V_\text{far}(r, \theta) 
    \end{pmatrix}
\end{align}
Notice that in Eq.~\eqref{Relation_dipolar_appendix} we have made use of the trigonometrical identity $(1 + \cos^2 \theta)^2 - 4 \cos^2 \theta =  \sin^4 \theta$. Finally, if we substitute $m = p$, which is valid for both an incident circularly polarized plane-wave and a circularly polarized focused Gaussian beam, we arrive at Eq.~\eqref{Relation_dipolar}.
\clearpage

\clearpage
\newpage

\bibliography{Bib_tesis}

\end{document}